\documentstyle[sprocl,psfig]{article}

\def\al{\alpha}

\def\be{\begin{equation}}
\def\ee{\end{equation}}
\def\bea{\begin{eqnarray}}
\def\eea{\end{eqnarray}}

\def\al{\alpha}
\def\be{\beta}
\def\ga{\gamma}
\def\de{\delta}

\def\th{\theta}

\def\ka{\kappa}

\def\si{\sigma}

\def\ps{\psi}

\def\De{\Delta}

\def\fr#1#2{{{#1} \over {#2}}}
\def\frac#1#2{{\textstyle{{#1}\over {#2}}}}

\def\ket#1{|{#1}\rangle}

\def\half{{\textstyle{1\over 2}}}
\def\lsim{\mathrel{\rlap{\lower4pt\hbox{\hskip1pt$\sim$}}
    \raise1pt\hbox{$<$}}}
\def\gsim{\mathrel{\rlap{\lower4pt\hbox{\hskip1pt$\sim$}}
    \raise1pt\hbox{$>$}}}
\def\sqr#1#2{{\vcenter{\vbox{\hrule height.#2pt
         \hbox{\vrule width.#2pt height#1pt \kern#1pt
         \vrule width.#2pt}
         \hrule height.#2pt}}}}

\def\hydrogen{H}
\def\antihydrogen{$\overline{\rm{H}}$}
\newcommand{\beq}{\begin{equation}}
\newcommand{\eeq}{\end{equation}}
\newcommand{\rf}[1]{(\ref{#1})}

\begin{document}

\begin{flushright}
IUHET 401\\
January 1999
\end{flushright}
\vskip 0.5 cm

\title{CPT AND LORENTZ SYMMETRY IN HYDROGEN AND ANTIHYDROGEN\footnote
{Talk at the Meeting on CPT and Lorentz Symmetry,
Indiana University, Bloomington, November 1998}}

\author{N.E. RUSSELL}

\address{Department of Physics, Indiana University, Bloomington, IN 47405, USA
\\E-Mail: nerussel@indiana.edu}


\maketitle
\abstracts{Possibilities for observing
signals of CPT and Lorentz violation 
in the spectroscopy of hydrogen and antihydrogen
are considered.
We show that transitions between the $c$ and $d$ hyperfine 
sublevels in the 1S state 
can exhibit theoretically detectable effects
that would be unsuppressed by powers of the 
fine-structure constant.
This transition may therefore offer some 
advantages over 1S-2S two-photon spectroscopy.
}

\section{Introduction}
The recent production and observation of antihydrogen (\antihydrogen) 
\cite{oelert,mandel}
opens new possibilities for 
precision tests of CPT symmetry.
The two-photon 1S-2S transition frequency
has been measured 
to $3.4$ parts in $10^{14}$
in an atomic beam of hydrogen (\hydrogen) 
\cite{hansch}
and to about one part in $10^{12}$
in trapped \hydrogen.\cite{cesar} 
It is hoped that
an eventual measurement of the line center 
to about $1$ mHz,
corresponding to a resolution of one part in $10^{18}$,
would be possible.\cite{hanschICAP} 
If such precisions could also be achieved 
in the spectroscopy of \antihydrogen,
comparisons of corresponding frequencies in
\hydrogen\ and \antihydrogen\
could yield stringent tests of CPT symmetry.
Current proposals for 
\antihydrogen\ spectroscopy
involve both beam and trapped-atom 
techniques,\cite{mandel2,gab2} 
and are faced with a number of 
outstanding challenges
including
the issue of achieving these precisions
in {\em trapped} \hydrogen\ and 
\antihydrogen.\cite{ce} 
We consider the 
theoretical prospects for placing appropriate bounds 
on CPT and Lorentz violation
in experiments involving the 
spectroscopy of free 
or magnetically trapped \hydrogen\ and \antihydrogen.

All local Lorentz-invariant 
quantum field theories of point 
particles,
including the standard model 
and quantum electrodynamics (QED),
are invariant under the discrete symmetry 
CPT.\cite{cpt}
Attempts to produce a fundamental theory 
involving gravity often involve 
string theory
and the spontaneous breaking of these symmetries\cite{kps}
and, 
in these investigations, 
the  status of CPT symmetry is far less clear.
Observable effects of CPT breaking 
are already known to be small,
and so it is reasonable to assume they 
would be suppressed by at least one power 
of the low-energy scale to Planck scale ratio.
Thus, 
their detection could occur 
only in extremely sensitive experiments.

In this proceedings,
we show that effects of this type
can appear in \hydrogen\ and \antihydrogen\ spectra
at zeroth order in the fine-structure constant. 
In addition,
these effects are theoretically detectable not only  
in 1S-2S lines but also in hyperfine transitions. 

The framework of our analysis is 
an extension of the standard model and QED\cite{ck}
that includes spontaneous CPT and Lorentz
breaking at a more fundamental level. 
Desirable features of this microscopic theory 
appear to include energy-momentum conservation,
gauge invariance,
renormalizability,
and microcausality.\cite{ck} 
Analyses in the context of this theoretical framework
have been done for 
photon properties,\cite{ck}
neutral-meson experiments,\cite{kps,ckpv,expt,ak}
Penning-trap tests,\cite{bkr}
and baryogenesis.\cite{bckp} 

\section{Free \hydrogen\ and \antihydrogen}
We first consider
the spectra of \it free \rm 
\hydrogen\ and \antihydrogen.
For \hydrogen,
the electron of mass $m_e$
and charge $q = -|e|$ 
in the proton Coulomb potential 
$A^\mu = (|e|/4 \pi r, 0)$
is described by a modified Dirac equation
arising from 
the standard-model extension.
Taking $i D_\mu \equiv i \partial_\mu - q A_\mu$,
the four-component electron field $\ps$ 
satisfies
\beq
\left( i \ga^\mu D_\mu - m_e - a_\mu^e \ga^\mu
- b_\mu^e \ga_5 \ga^\mu 
- \half H_{\mu \nu}^e \si^{\mu \nu} 
+ i c_{\mu \nu}^e \ga^\mu D^\nu 
+ i d_{\mu \nu}^e \ga_5 \ga^\mu D^\nu \right) \ps = 0
\label{dirac}
\eeq
in units with $\hbar = c = 1$.
CPT is violated by 
the two terms involving the couplings
$a_\mu^e$ and $b_\mu^e$,
while CPT is preserved by
the three terms involving 
$H_{\mu \nu}^e$, $c_{\mu \nu}^e$, and $d_{\mu \nu}^e$.
Lorentz invariance is broken by
all five couplings, 
which are assumed to be small.\cite{ck}
Free protons are also described by
a modified Dirac equation\cite{bkr}
with corresponding couplings
$a_\mu^p$, $b_\mu^p$, $H_{\mu \nu}^p$, $c_{\mu \nu}^p$, 
and $d_{\mu \nu}^p$.
It is possible to eliminate 
various combinations of these quantities
through suitable field redefinitions.
In the following,
we keep all couplings,
thus showing explicitly that these expressions
are unobservable.\cite{ck}

Observable effects in the spectra of 
free \hydrogen\ and \antihydrogen\
can be studied using 
perturbative calculations
in the context of relativistic quantum mechanics. 
In this calculation,
the unperturbed hamiltonians
and their eigenfunctions are
identical for \hydrogen\ and \antihydrogen. 
In addition, 
all perturbative effects
from conventional quantum electrodynamics
are also the same in both systems. 
However,
the perturbations arising from 
the CPT- and Lorentz-breaking couplings
for the electron in \hydrogen\
can differ from those 
for the positron in \antihydrogen. 
These perturbations are obtained
from Eq.\ \rf{dirac}
by a standard method involving charge conjugation
(for \antihydrogen)
and field redefinitions.\cite{bkr}
Similarly,
additional energy perturbations
are generated by 
the CPT- and Lorentz-breaking couplings 
for the proton and antiproton,
and can be obtained to leading order via
relativistic two-fermion techniques.\cite{D2}

Let the (uncoupled) electronic and nuclear angular momenta
be denoted by
$J=1/2$ and $I=1/2$
respectively,
with third components $m_J$, $m_I$. 
Using a perturbative calculation,
the energy corrections for the basis states $\ket{m_J,m_I}$
can be found.
For protons or antiprotons,
we find that the leading-order energy corrections 
for spin eigenstates 
have the same mathematical form as those 
for electrons or positrons,
except for the replacement of superscripts $e$ with $p$
on the CPT- and Lorentz-violating couplings. 

In \hydrogen, 
we find that the 
leading-order energy shifts in
the 1S level are identical to those 
in the 2S level.
Taking $m_p$ for the proton mass, 
the shifts are 
\bea
\De E^{H} (m_J, m_I)
& \approx &
(a_0^e + a_0^p - c_{00}^e m_e - c_{00}^p m_p)
\cr
&&
+ (-b_3^e + d_{30}^e m_e + H_{12}^e) {m_J}/{|m_J|} 
\cr
&&
+ (-b_3^p + d_{30}^p m_p + H_{12}^p) {m_I}/{|m_I|} ~ .
\label{EHJI}
\eea
Similarly,
for \antihydrogen,
the leading-order energy shifts 
$\De E^{ \overline{H}}$
in the 1S levels 
are identical to those in 
the 2S levels,
and are
given by the expression \rf{EHJI}
with the substitutions
$a_\mu^e \rightarrow - a_\mu^e$,
$d_{\mu \nu}^e \rightarrow - d_{\mu \nu}^e$,
$H_{\mu \nu}^e \rightarrow - H_{\mu \nu}^e$;
$a_\mu^p \rightarrow - a_\mu^p$,
$d_{\mu \nu}^p \rightarrow - d_{\mu \nu}^p$,
$H_{\mu \nu}^p \rightarrow - H_{\mu \nu}^p$. 
We note that because Eq.~(\ref{EHJI}) contains 
spatial components of the couplings,
it would be necessary to take into account the geometry 
when comparing results from different 
experiments.
For example,
measurements taken at different times of the day
would be sensitive to different projections 
of the couplings due to the rotation of the Earth.

The electron and proton spins in \hydrogen\
are coupled by the hyperfine interaction,
and this is also the case for the 
positron and antiproton spins
in \antihydrogen. 
The total angular momentum $F$
must be considered,
and the appropriate basis states become 
linear combinations $\ket{F,m_F}$
of the $\ket{m_J,m_I}$ states. 
The allowed two-photon 1S-2S transitions
satisfy the selection rules
$\De F = 0$ and $\De m_F = 0$.\cite{cagnac}
There are thus four allowed transitions 
for both \hydrogen\ and \antihydrogen,
those for which the spins remain unchanged.
However,
no leading-order effects appear in the frequencies
of any of these transitions,
because according to Eq.\ \rf{EHJI}
the 1S and 2S states with identical spin configurations
have identical leading-order energy shifts.
Thus in the present theoretical context,
there are no signals of 
Lorentz or CPT violation in 
free \hydrogen\ or in free 
\antihydrogen\ at leading-order in 1S-2S 
spectroscopy.
This agrees with results found previously\cite{bkr}
for the Penning trap,
showing that observable CPT-violating effects must also involve
CT violation and a spin-flip.

To overcome this limitation,
one could consider 
the dominant subleading energy-level shifts
involving the CPT- and Lorentz-breaking couplings 
in free \hydrogen\ and \antihydrogen.
These would be hard to detect because they
arise as relativistic corrections of order $\al^2$. 
They do, however,  
differ for some of the 1S and 2S levels
and therefore observable effects
could in principle occur.
An example is the term proportional to 
$b_3^e$ in Eq.\ \rf{dirac},
which produces a frequency shift in the 
$m_F = 1 \rightarrow m_{F^\prime} = 1$ line
relative to the $m_F = 0 \rightarrow m_{F^\prime} = 0$ line
(which remains unshifted),
given by 
\beq
\de \nu^H_{1S-2S} \approx - \al^2 b_3^e / 8 \pi 
\quad .
\eeq
A similar suppression by a factor 
at least of order
$\al^2 \simeq 5\times 10^{-5}$
would occur in 
the proton-antiproton corrections.
As a result of these suppressions,
Penning-trap $g-2$ experiments
are likely to be more sensitive to 
some of the CPT- and Lorentz-violating quantities 
than experiments involving 1S-2S spectroscopy
in free \hydrogen\ and \antihydrogen.
In fact,
the estimated attainable bound\cite{bkr}
on $b_3^e$ 
obtained 
with existing technology
in anomaly-frequency comparisons
with electron-positron Penning-trap experiments 
would suffice to place a bound 
of $\de \nu^H_{1S-2S} \lsim 5$ $\mu$Hz
on observable shifts of the 1S-2S frequency 
in free \hydrogen\ from the electron-positron sector. 
This is beyond the resolution of 1S-2S spectroscopy. 
For the proton-antiproton quantities 
in the standard-model extension,
experiments have not yet been performed,
but bounds attainable would 
also yield tighter constraints on these
parameters than would be possible in 1S-2S spectroscopy. 

It is relevant to ask why 
$g-2$ experiments 
are potentially more sensitive to 
observable effects than
comparisons of 1S-2S
transitions in free \hydrogen\ and \antihydrogen. 
This is surprising because
the conventional figure of merit for CPT breaking 
in electron-positron $g-2$ experiments,\cite{pdg}
\beq
r_g = |g_{e^-} - g_{e^+}|/g_{\rm av} \lsim 2 
\times 10^{-12}
\quad ,
\eeq
is six orders of magnitude weaker
than the idealized resolution 
of the 1S-2S line,
$\De \nu_{1S-2S}/\nu_{1S-2S} \simeq 10^{-18}$.
However,
the figure of merit $r_g$ in Penning-trap 
$g-2$ experiments is inappropriate
in the present theoretical context.\cite{bkr} 
The point is that the 
experimental sensitivity 
to CPT- and Lorentz-violating effects
is determined by 
the absolute frequency resolution 
for unsuppressed transitions.
The idealized 1S-2S line-center resolution
is about 1 mHz,
which would appear to be better than 
the 1 Hz absolute frequency resolution 
in $g-2$ measurements.
However,
$g-2$ experiments are 
directly sensitive to $b_3^e$
because they involve spin-flip transitions,
whereas the 1S-2S transitions 
in free \hydrogen\ or \antihydrogen\ 
are sensitive only to the suppressed combination 
$\al^2 b_3^e/8\pi$. 
As a result,
the bound on $b_3^e$ 
from electron-positron $g-2$ experiments
is thus about two orders of magnitude sharper 
than that from 1S-2S comparisons.

In addition to the 1S-2S transition,
there are certainly others
available in \hydrogen\ and \antihydrogen.
The above discussion suggests 
that transitions
between states with different spin configurations
might yield tighter bounds. 
Such experiments would require external fields
to select particular spin states.

\section{Trapped \hydrogen\ and \antihydrogen}
We next consider spectroscopy of 
\hydrogen\ or \antihydrogen\ 
in the presence of a uniform magnetic field.
A way to do this is by confining the particles 
in a magnetic trap
such as an Ioffe-Pritchard trap,\cite{ip} 
and imposing an axial bias magnetic field.
The situation is directly relevant to 
proposed experiments.\cite{gab2} 
In the following,
we denote each of the 1S and 2S hyperfine Zeeman levels 
in order of increasing energy 
in a magnetic field $B$ by  
$\ket{a}_n$, $\ket{b}_n$, $\ket{c}_n$, $\ket{d}_n$,
with $n=1$ or $2$, 
for both \hydrogen\ and \antihydrogen. 
In the case of \hydrogen,
the four states expressed in terms of the 
basis states $\ket{m_J,m_I}$ are 
\bea
\ket{d}_n &=& \ket{\half, \half} \quad, 
\nonumber \\
\ket{c}_n &=& \sin \th_n \ket{-\half,\half} +
\cos \th_n \ket{\half,-\half}
\quad ,
\nonumber\\
\ket{b}_n &=& \ket{-\half, -\half} \quad, 
\nonumber \\
\ket{a}_n &=& \cos \th_n \ket{-\half,\half} -
\sin \th_n \ket{\half,-\half}
\quad . 
\label{a}
\eea
The mixing angles $\th_n$
are functions of the magnetic field,
and are different for the 1S and 2S states:
\beq
\tan 2 \th_n \approx \fr{(51 {\rm ~mT})}{n^3B}
\quad . 
\eeq
The states 
$\ket{c}_1$ and $\ket{d}_1$
are low-field seekers, 
and in principle remain confined 
near the magnetic-field minimum of the trap.
However, 
a population loss occurs 
due to spin-exchange collisions 
$\ket{c}_1 + \ket{c}_1 \rightarrow \ket{b}_1 + \ket{d}_1$ 
of the $\ket{c}_1$ states over time,
so that primarily $\ket{d}_1$ states
are confined.

A transition that would seem natural to consider
is that between the unmixed-spin states 
$\ket{d}_1$ and $\ket{d}_2$ 
because it is field independent
for practical values of the magnetic field. 
The idea would be to compare
the frequency $\nu^H_d$ 
for the 1S-2S transition $\ket{d}_1 \rightarrow \ket{d}_2$ 
in \hydrogen\ 
with the frequency $\nu^{\overline{H}}_d$ 
for the corresponding spectroscopic line in \antihydrogen. 
But,
in \hydrogen\ the spin configurations of the 
$\ket{d}_1$ and $\ket{d}_2$ states are the same,
so any shifts occurring are again suppressed.
The same is true for \antihydrogen,
and so we find 
\beq
\de \nu^H_d = \de \nu^{\overline{H}}_d \simeq 0
\eeq
at leading order.

\begin{figure}[t]
\hspace*{1.4cm}
\centerline{
\psfig{figure=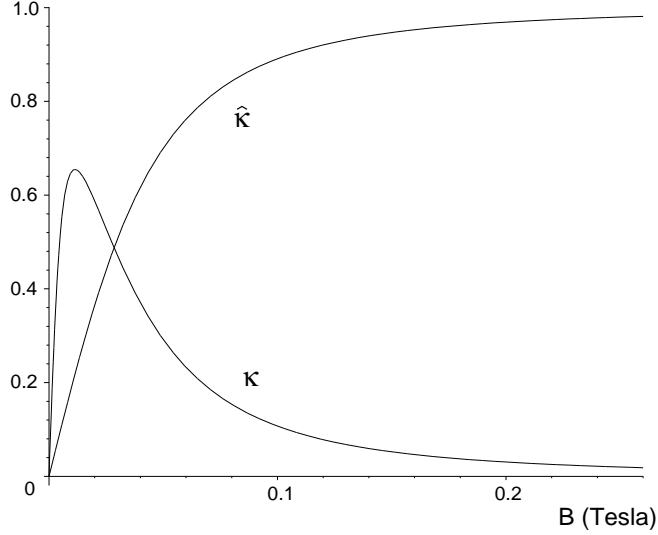,height=3in,width=3in}}
\caption{The dimensionless functions $\ka$ and $\hat{\ka}$.
For $\ka$, the maximum value of approximately 0.67 occurs 
at about 0.011~T.
The function $\hat{\ka}$ increases to within 
about two percent of its asymptotic value (one)
as the magnetic field is increased from zero 
to $0.25$ Tesla.
\label{kapfunc}}
\end{figure}

Another transition of theoretical interest
would be the 1S-2S transition
$\ket{c}_1 \rightarrow \ket{c}_2$ in \hydrogen\ 
and the analogous \antihydrogen\ transition. 
The point would be to exploit 
the spin mixing of these states in a nonzero magnetic field. 
An unsuppressed frequency shift 
would arise because the hyperfine splitting 
depends on $n$,
thus producing a spin difference 
between the 1S and 2S levels
in this 1S-2S transition
between $\ket{c}_1$ and $\ket{c}_2$:
\beq
\de \nu_c^H \approx
-\ka (b_3^e - b_3^p - d_{30}^e m_e 
+ d_{30}^p m_p - H_{12}^e + H_{12}^p)/2\pi ~.
\label{nucH}
\eeq
In this expression,
$\ka$ is a spin-mixing function
given by
\beq
\ka\equiv \cos 2\th_2 - \cos 2\th_1
\quad .
\eeq
This function is always less than one,
so to avoid losing sensitivity
the optimal situation would involve the largest
possible value.
This maximum is $\ka \simeq 0.67$
and occurs at $B \simeq 0.011$~T,
as illustrated in Figure~\ref{kapfunc}.

The corresponding 1S-2S shift 
$\de \nu_c^{\overline{H}}$ for \antihydrogen\ 
in the same magnetic field can also be found. 
Relative to a fixed magnetic field,
the hyperfine states in \antihydrogen\ 
have opposite positron and antiproton spins 
compared to the electron and proton spins in \hydrogen.
As a result,
the expression for 
$\de \nu_c^{\overline{H}}$ is 
identical to that for $\de \nu_c^H$ in Eq.\ \rf{nucH}
except that the signs of $b_3^e$ and $b_3^p$ are changed. 
The frequencies $\nu_c^H$ and $\nu_c^{\overline H}$
depend on spatial components of Lorentz-violating couplings
and would therefore vary diurnally 
in the comoving Earth frame. 
Another effect would be an istantaneous difference
\beq
\De \nu_{1S-2S,c} \equiv \nu_c^H 
- \nu_c^{\overline{H}} \approx - \ka (b_3^e - b_3^p)/\pi
\label{delcc}
\eeq
for measurements made in the same magnetic trapping fields. 

The transition $\ket{c}_1 \rightarrow \ket{c}_2$ 
when compared with 
the transition $\ket{d}_1 \rightarrow \ket{d}_2$
is theoretically more sensitive to 
CPT and Lorentz violation 
by a factor of order
$4/\al^2 \simeq 10^5$. 
However,
the 1S-2S transition
$\ket{c}_1 \rightarrow \ket{c}_2$
in \hydrogen\ and \antihydrogen\ 
depends on the magnetic field,
and the resultant Zeeman broadening 
due to the inhomogeneous trapping fields
would have to be overcome.
Even at a temperature of $100 \mu$K, 
the transition 
in both \hydrogen\ and \antihydrogen\
would be broadened to over 1 MHz 
for $B\simeq 10$ mT.
This would severely hinder
the experimental attainment of 
resolutions on the order of the natural line width.

\begin{figure}[t]
\hspace*{3cm}
\centerline{
\psfig{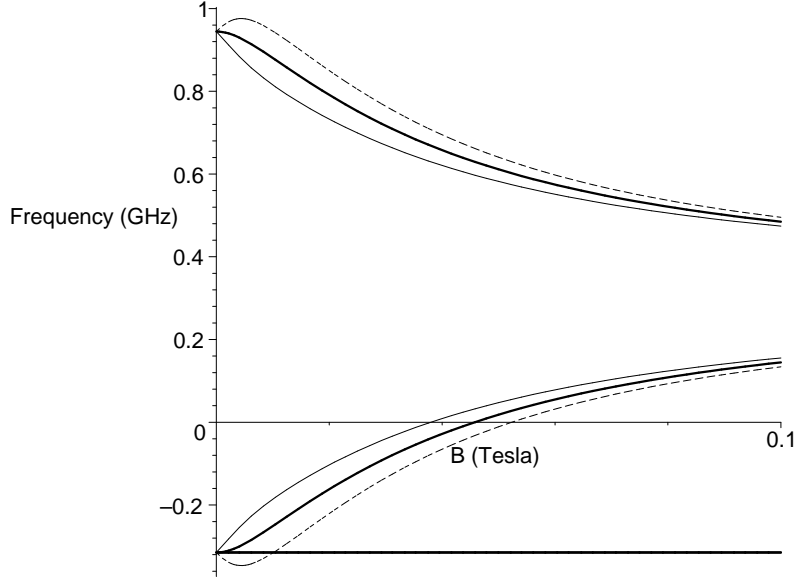}}
\caption{Conventional and perturbed frequencies 
for the 1S-2S transition as a function of magnetic field.
The vertical scale is the shift in the 
usual Bohr-model 1S-2S frequency of 
about $2.5\times 10^{15}$~Hz.
The bold lines are for the conventional frequencies,
the fainter solid line is for the perturbed hydrogen transition 
frequencies, and the dashed line is for 
the perturbed antihydrogen frequencies.
We have taken $b_3^p>0$, with all other couplings zero.
The upper set of three lines represents the 
$\ket{a}_1 \rightarrow \ket{a}_2$ 
transition,
and the lower set the 
$\ket{c}_1 \rightarrow \ket{c}_2$ 
case.
The single straight line is for the 
$\ket{b}_1 \rightarrow \ket{b}_2$ 
and 
$\ket{d}_1 \rightarrow \ket{d}_2$ cases,
showing how these transitions are 
field independent and 
unperturbed by the $b_3^p$ coupling.
\label{1s2sfig}}
\end{figure}

Figure \ref{1s2sfig} illustrates one case for the conventional and 
perturbed frequencies in the four 1S-2S transitions.
In this figure, $b_3^p>0$ and all the other couplings are zero.

\section{Hyperfine Transitions}
We now consider the possibilities for 
spectroscopy of the hyperfine 1S levels.
Motivated by the fact that 
transitions between the $F = 0$ and $F^\prime = 1$
hyperfine states can be measured with accuracies
better than $1$ mHz in a hydrogen maser,\cite{ramsey}
hyperfine transitions in masers and
in trapped \hydrogen\ and \antihydrogen\
are worth considering
for tests of CPT and Lorentz symmetry.

\begin{figure}[t]
\vskip 2cm
\hspace*{2cm}
\centerline{
\psfig{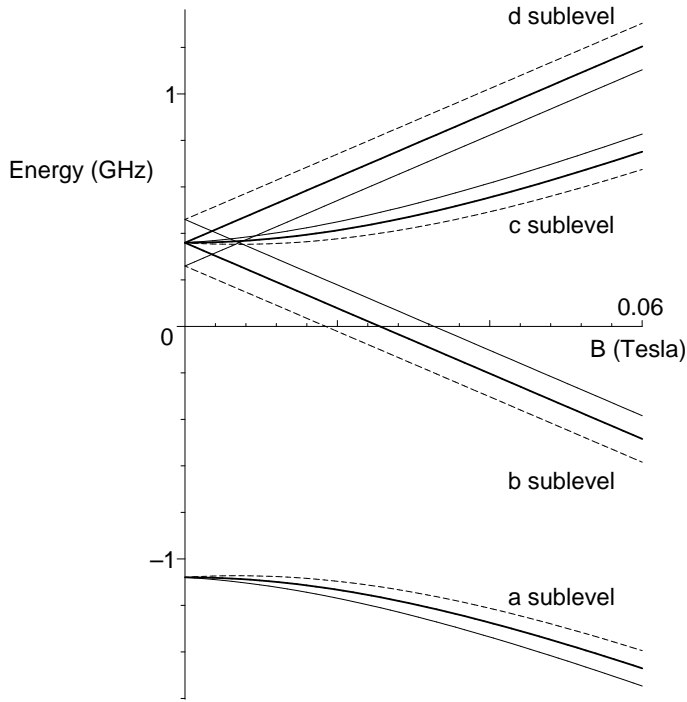}}
\caption{Hyperfine levels for 1S states
versus magnetic field.
The vertical axis represents the 
shift in energy (in frequency units) relative to the 
usual Bohr-model $n=1$ energy of $-13.6$~eV.
The bold solid line is for the unperturbed case,
the finer solid line and the dashed lines are 
for hydrogen and antihydrogen respectively.
We have taken $b_3^p>0$ 
and all other couplings zero.
\label{fig1}}
\end{figure}

The energy levels of all four hyperfine states 
in the ground state of hydrogen 
are shifted due to 
CPT- and Lorentz-violating effects. 
All the shifts contain an identical contribution 
$a_0^e + a_0^p -c_{00}^e m_e -c_{00}^p m_p$
that leaves energy differences unaffected. 
The remaining spin-dependent terms 
are 
\bea
\De E_a^H &\simeq&  
\hat\ka (b_3^e - b_3^p - d_{30}^e m_e 
+ d_{30}^p m_p - H_{12}^e + H_{12}^p)
\quad ,
\nonumber\\
\De E_b^H &\simeq& 
b_3^e + b_3^p - d_{30}^e m_e 
- d_{30}^p m_p - H_{12}^e - H_{12}^p
\quad ,
\nonumber\\
\De E_c^H &\simeq& -\De E_a^H
\quad , \qquad
\De E_d^H \simeq - \De E_b^H
\quad ,
\label{abcd}
\eea
where $\hat\ka \equiv \cos2 \th_1$. 
If there is no magnetic field, then  $\hat\ka =0$
and the energies of $\ket{a}_1$ and $\ket{c}_1$ are unshifted. 
However, 
equal and opposite energy shifts occur for 
$\ket{b}_1$ and $\ket{d}_1$.
The degeneracy of the three $F=1$ ground-state
hyperfine levels is therefore removed even for 
$B=0$.\footnote{No conflict with Kramer's theorem occurs
in the breaking of the 
$\ket{b}$-$\ket{d}$ degeneracy at zero field,
because the Lorentz-violating coefficients 
in Eq.~\rf{abcd} break time-reversal symmetry.
A possible method of detecting the splitting might 
involve looking directly for a difference frequency.}
For instance,
the 
$\ket{d}_1 \rightarrow \ket{a}_1$
and $\ket{b}_1 \rightarrow \ket{a}_1$ 
transitions differ in their frequencies 
by the unsuppressed and diurnally varying 
quantity
\beq
|\De \nu_{d-b}^H| \approx
|b_3^e + b_3^p - d_{30}^e m_e - d_{30}^p m_p
- H_{12}^e - H_{12}^p|/\pi
\quad . 
\eeq
In the presence of a magnetic field,
all four hyperfine Zeeman energy levels are shifted. 
For the $\ket{a}_1$ and $\ket{c}_1$ states,
the spin-mixing function $\hat\ka$
controls the shifts.
As $B$ increases from zero,
$\hat\ka$ increases,
attaining $\hat\ka \simeq 1$ when $B \simeq 0.3$ T. 
The function $\hat\ka$ is illustrated in 
Fig.~\ref{kapfunc}.
The shifts in the energy levels as given in 
Eq.~\rf{abcd} are partially illustrated in 
Figure~\ref{fig1}.

The usual \hydrogen\ maser employs 
a small ($B \lsim 10^{-6}$ T) magnetic field
and works with the
field-independent $\si$ transition 
$\ket{c}_1 \rightarrow \ket{a}_1$.
The leading-order effects from 
CPT and Lorentz violation in high-precision
measurements of this line 
$\ket{c}_1 \rightarrow \ket{a}_1$ are suppressed,
because for this situation $\hat\ka \lsim 10^{-4}$.
However,
a shift $\De \nu_{d-b}^H$ does occur in 
the frequency difference between the 
field-dependent transitions $\ket{d}_1 \rightarrow \ket{a}_1$
and $\ket{b}_1 \rightarrow \ket{a}_1$
relative to the conventional value,
and the associated diurnal variations 
would provide an unsuppressed signal 
of CPT and Lorentz violation. 
The resolution of this difference 
would be reduced by 
broadening due to field inhomogeneities.
In addition, it would be necessary to 
distinguish it from possible backgrounds 
due to residual Zeeman splittings. 

The direct comparison of transitions
between hyperfine Zeeman levels in \hydrogen\ and \antihydrogen\
could address the issue of background splittings.
Moreover,
the magnetic-field dependence of the frequency
could be eliminated to first order 
by working at an appropriate value of the field.
One option might be to consider 
high-resolution spectroscopy
at the field-independent transition point $B \simeq 0.65$~T
on the $\ket{d}_1 \rightarrow \ket{c}_1$ transition
in trapped \hydrogen\ and \antihydrogen.
Experimental hurdles would include
Doppler broadening and 
potentially larger field inhomogeneities
due to the relatively high bias field.
Obtaining  frequency resolutions of order 1 mHz
would be a challenge,
requiring cooling to temperatures of order 100 $\mu$K
with a good signal-to-noise ratio
and a stiff box shape for the trapping potential.

At this bias-field strength,
the electron and proton spins 
in state $\ket{c}_1$ 
interact more strongly with the field 
than with each other
and are highly polarized with
$m_J = 1/2$ and $m_I = - 1/2$. 
Thus, the transition $\ket{d}_1 \rightarrow \ket{c}_1$ 
is in essence a proton spin-flip.
For this transition,
we obtain frequency shifts 
\bea
\de \nu_{c \rightarrow d}^H  &\approx&
(-b_3^p + d_{30}^p m_p + H_{12}^p)/\pi 
\quad ,
\nonumber\\
\de \nu_{c \rightarrow d}^{\overline{H}} &\approx& 
(b_3^p + d_{30}^p m_p + H_{12}^p)/\pi
\eea
for \hydrogen\ and 
\antihydrogen\ respectively.
One way to detect such terms 
would be to search for diurnal variations
in the frequencies
$\nu_{c \rightarrow d}^H$ 
and
$\nu_{c \rightarrow d}^{\overline{H}}$.
Another possibility would be to consider 
their instantaneous difference,
\beq
\De \nu_{c \rightarrow d} \equiv 
\nu_{c \rightarrow d}^H - \nu_{c \rightarrow d}^{\overline{H}}
\approx - 2 b_3^p / \pi
\quad .
\label{nudiff}
\eeq
This difference 
could provide a direct, clean, and sharp test 
of the CPT-violating coupling $b_3^p$ for the 
proton.

We can introduce
dimensionless figures of merit 
appropriate for experiments investigating 
various direct and diurnal-variation signals.
This is done in analogy with definitions 
made for similar tests in Penning traps.\cite{bkr} 
As an example,
a figure of merit for the signal in Eq.\ \rf{nudiff}
could be chosen as
\bea
r^H_{rf,c \rightarrow d} & \equiv &
{|({\cal E}_{1,d}^H - {\cal E}_{1,c}^H)
- ({\cal E}_{1,d}^{\overline{H}} - {\cal E}_{1,c}^{\overline{H}})|}/
{{\cal E}_{1,{\rm av}}^H}
\nonumber \\
&\approx &
2\pi |\De \nu_{c \rightarrow d}| /m_H
\quad .
\label{rrf}
\eea
Here, ${\cal E}_{1,d}^H$, ${\cal E}_{1,c}^H$ and the
corresponding quantities for \antihydrogen\ 
are relativistic energies in ground-state hyperfine levels,
and $m_H$ is the atomic mass of \hydrogen. 
If, for example, 
a frequency resolution of 1 mHz were attained,
this would correspond to an upper bound 
of about 
$r^H_{rf,c \rightarrow d} \lsim 5 \times 10^{-27}$. 
The CPT- and Lorentz-violating coupling $b_3^p$ 
would then be limited to $|b_3^p| \lsim 10^{-18}$~eV.
This is about three orders of magnitude better
than estimated attainable bounds\cite{bkr}
from $g-2$ experiments in Penning traps
and more than four orders of magnitude better
than the limit attainable from 1S-2S 
transitions.
We also note that the frequency resolution 
of high-precision clock-comparison experiments,
which can also bound Lorentz violation,
lies below 1 $\mu$Hz.\cite{hd}
In these experiments,
leading-order bounds are obtained on $b_3^p$
in combination with other couplings.\cite{kla}
Since the nuclei involved are relatively complex,
the theoretical analysis prevents
$b_3^p$ from being isolated.

The experiments discussed here are sensitive 
only to spatial components of CPT-violating couplings.
A boost would be needed to be sensitive to 
timelike components such as 
$b_0^e$,
and would also 
enhance CPT- and Lorentz-violating
effects.\cite{ak}
This would be an advantage of the proposed experiments
\cite{mandel2}
measuring the fine structure and Lamb shift
with a relativistic beam of \antihydrogen.
Although they would probably have poorer resolutions
than the others discussed here,
constraints on $b_0^e$ and $b_0^p$ may be possible.

In conclusion,
we have shown that 1S-2S transitions
involving the mixed-spin $\ket{c}$ states
as well as the 
spin-flip 
$\ket{d}_1 \rightarrow \ket{c}_1$ 
hyperfine transition
could give rise to signals of
Lorentz and CPT violation 
in magnetically confined 
\hydrogen\ or \antihydrogen\ atoms.
These signals would not be suppressed by powers
of the fine-structure constant. 
They would indicate observable and 
qualitatively new physics originating at the Planck scale. 

\section*{Acknowledgments}
I thank Robert Bluhm and Alan Kosteleck\'y,
who collaborated on this work.
Partial support was provided by the U.S.\ D.O.E.\
under grant number DE-FG02-91ER40661 and by the N.S.F.\ 
under grant number PHY-9503756.


\section*{References}
\begin{thebibliography}{99}

\bibitem{oelert}
G.\ Baur 
{\it et al.},
Phys.\ Lett.\ B {\bf 368} (1996) 251.

\bibitem{mandel}
G.\ Blanford 
{\it et al.},
Phys.\ Rev.\ Lett. {\bf 80} (1998) 3037.

\bibitem{hansch}
T.\ Udem 
{\it et al.},
Phys.\ Rev.\ Lett.\ {\bf 79} (1997) 2646.

\bibitem{cesar}
C.L.\ Cesar 
{\it et al.},
Phys.\ Rev.\ Lett.\ {\bf 77} (1996) 255.

\bibitem{hanschICAP}
See, for example,
T.W.\ H\"ansch, 
in D.J.\ Wineland, C.E.\ Wieman,
and S.J.\ Smith, eds.,
{\it Atomic Physics 14}
(A.I.P.\, New York, 1995).

\bibitem{mandel2}
G.\ Blanford
{\it et al.},
Phys.\ Rev. D {\bf 57} (1998) 6649.

\bibitem{gab2}
B.\ Brown
{\it et al.},
Nucl.\ Phys.\ B (Proc. Suppl.) {\bf 56A} (1997) 326;
M.H.\ Holzscheiter
{\it et al.},
{\it ibid.}, 336.

\bibitem{ce}
See, for example,
M.H.\ Holzscheiter and M.\ Charlton,
Rep.\ Prog.\ Phys.\ {\bf 62} (1999) 1;
M.\ Charlton
{\it et al.},
Phys.\ Rep.\ {\bf 241} (1994) 65;
J.\ Eades, ed.,
{\it Antihydrogen}
(J.C.\ Baltzer, Geneva, 1993).

\bibitem{cpt}
For a discussion of the CPT theorem,
see, for example,
R.F.\ Streater and A.S.\ Wightman,
{\em PCT, Spin and Statistics and All That,}
Benjamin Cummings, Reading, 1964.

\bibitem{kps}
V.A.\ Kosteleck\'y and S.\ Samuel,
Phys.\ Rev.\ Lett.\ {\bf 63} (1989) 224;
{\it ibid.},
{\bf 66} (1991) 1811;
Phys.\ Rev. D {\bf 39} (1989) 683;
{\it ibid.},
{\bf 40} (1989) 1886;
V.A.\ Kosteleck\'y and R.\ Potting,
Nucl.\ Phys.\ B {\bf 359} (1991) 545;
Phys.\ Lett.\ B {\bf 381} (1996) 89.

\bibitem{ck}
D.\ Colladay and V.A.\ Kosteleck\'y,
Phys.\ Rev.\ D {\bf 55} (1997) 6760;
{\it ibid.}, {\bf 58} (1998) 116002.

\bibitem{ckpv}
V.A.\ Kosteleck\'y and R.\ Potting,
in D.B.\ Cline, ed.,
{\it Gamma Ray--Neutrino Cosmology and Planck Scale Physics} \rm
(World Scientific, Singapore, 1993)
(hep-th/9211116);
Phys.\ Rev.\ D {\bf 51} (1995) 3923;
D.\ Colladay and V. A. Kosteleck\'y,
Phys.\ Lett.\ B {\bf 344} (1995) 259;
Phys.\ Rev.\ D {\bf 52} (1995) 6224;
V.A.\ Kosteleck\'y and R.\ Van Kooten,
Phys.\ Rev. D {\bf 54} (1996) 5585.

\bibitem{expt}
OPAL Collaboration, 
R.\ Ackerstaff
{\it et al.},
Z.\ Phys. C {\bf 76} (1997) 401;
DELPHI Collaboration,
M.\ Feindt
{\it et al.},
preprint DELPHI 97-98 CONF 80 (July 1997).

\bibitem{ak}
V.A.\ Kosteleck\'y,
Phys.\ Rev.\ Lett.\ {\bf 80} (1998) 1818.

\bibitem{bkr}
R.\ Bluhm, V.A.\ Kosteleck\'y and N.\ Russell,
Phys.\ Rev.\ Lett.\ {\bf 79} (1997) 1432;
Phys.\ Rev.\ D {\bf 57} (1998) 3932.

\bibitem{bckp}
O.\ Bertolami
{\it et al.},
Phys.\ Lett.\ B {\bf 395} (1997) 178.

\bibitem{D2}
See, for example,
G. Breit,
Phys.\ Rev.\ {\bf 34} (1929) 553.

\bibitem{cagnac}
B.\ Cagnac, G.\ Grynberg, and F.\ Biraben,
J.\ Physique {\bf 34} (1973) 845.


\bibitem{pdg}
See, for example,
R.M.\ Barnett 
{\it et al.},
Review of Particle Properties,
Phys.\ Rev.\ D {\bf 54} (1996) 1.

\bibitem{ip}
Y.V.\ Gott, M.S.\ Ioffe, and V.G.\ Tel'kovskii,
Nucl.\ Fusion, Suppl.\ Pt.\ 3 (1962) 1045;
D.E.\ Pritchard,
Phys.\ Rev.\ Lett.\ {\bf 51} (1983) 1336.


\bibitem{ramsey}
N.F.\ Ramsey,
Physica Scripta {\bf T59} (1995) 323.


\bibitem{hd}
V.W.\ Hughes, H.G.\ Robinson, and V.\ Beltran-Lopez,
Phys.\ Rev.\ Lett.\ {\bf 4} (1960) 342;
R.W.P.\ Drever,
Philos.\ Mag.\ {\bf 6} (1961) 683;
J.D.\ Prestage 
{\it et al.},
Phys.\ Rev.\ Lett.\ {\bf 54} (1985) 2387;
S.K.\ Lamoreaux 
{\it et al.},
{\it ibid.}, 
{\bf 57} (1986) 3125;
T.E.\ Chupp
{\it et al.},
{\it ibid.}, 
{\bf 63} (1989) 1541;
C.J.\ Berglund
{\it et al.},
{\it ibid.}, 
{\bf 75} (1995) 1879.

\bibitem{kla}
V.A.\ Kosteleck\'y and C.D.\ Lane,
preprint IUHET 403 (1999).

\end{thebibliography}
\end{document}